\documentclass[revtex4,11pt]{article}
\usepackage[UKenglish]{babel}
\usepackage{amssymb}
\usepackage{mathrsfs}
\usepackage{mathtools}
\usepackage{dsfont}
\usepackage{amsmath,amsthm}
\usepackage{amsfonts}
\usepackage{enumerate, xspace}
\RequirePackage{hyperref}
\usepackage[bottom,first]{draftcopy}
\usepackage{changebar}
\usepackage{hyperref}
\usepackage{pdfsync}
\usepackage{graphicx}
\usepackage{verbatim}
\usepackage{a4wide}
\usepackage{pgfplots}
\usepackage{relsize}
\usepackage{bbm}
\usepackage{systeme}
\usepackage{mathtools}
\usepackage{tikz}
\usepackage{pgfplots}
\usepackage{graphicx}
\usepackage{adjustbox}
\usepackage{bytefield}
\usepackage{empheq}
\usepackage[labelfont=bf,margin=0.975cm,small]{caption}
\usepackage[toc,page]{appendix}
\usepackage[left=2.5cm,right=2.5cm]{geometry}

\usetikzlibrary{calc}
\usetikzlibrary{trees,snakes}
\usetikzlibrary{positioning}
\usepgfplotslibrary{fillbetween}
\usetikzlibrary{patterns}

\numberwithin{equation}{section}

\begin{document}

\begin{center}
{\bf\LARGE A Multispin Algorithm for the Kob-Andersen Stochastic Dynamics on Regular Lattices} \\
\end{center}
\begin{center}
{\Large Roberto Boccagna} \\ 
\vspace{2mm}
{Gran Sasso Science Institute, Viale F. Crispi 7, 67100 L'Aquila, Italy\\
{\hspace{-5.49cm}e-mail: \texttt{roberto.boccagna@gssi.it}}}
\end{center}

\vspace{3mm}

\abstract{
The aim of the paper is to propose an algorithm based on the Multispin Coding technique for the Kob-Andersen glassy dynamics. We first give motivations to speed up the numerical simulation in the context of spin glass models \cite{PAR}; after defining the Markovian dynamics as in \cite{KA} as well as the related interesting observables, we extend it to the more general framework of Random Regular Graphs, listing at the same time some known analytical results \cite{TONI}. The purpose of this work is a dual one; firstly, we describe how bitwise operators can be used to build up the algorithm by carefully exploiting the way data are stored on a computer. Since it was first introduced \cite{REBBI1,REBBI2}, this technique has been widely used to perform Monte Carlo simulations for Ising and Potts spin systems; however, it can be successfully adapted to more complex systems in which microscopic parameters may assume boolean values. Secondly, we introduce a random graph in which a characteristic parameter allows to tune the possible transition point. A consistent part is devoted to listing the numerical results obtained by running numerical simulations.

\section{Introduction}
\label{intro}
Glassy materials exhibit peculiar thermodynamical properties that mainly arise from the random nature of the microscopic interactions, which compete at the energetic level preventing the system to reach a stationary state \cite{PAR}. The dynamics is then typically constrained for a long time around metastable configurations characterized by disordered atomic dispositions in which clusters of correlated particles emerge. The size of these so called dynamical heterogeneities increases together with relaxation times to equilibrium when the temperature is lowered \cite{BER}. A key role is also played by the density $\rho$: when large enough, some particles may be trapped by their neighbors and remain blocked until the local configuration changes. This effect even contributes to the persistence of disordered configurations and then to the growth of the lifetime of the metastable states. 
\newline
In the last decades, a lot of mathematical models were proposed in order to mimic this effect rather than those induced by frustrated bonds. In most of these Kinetically Constrained Models (KCM), in which particles move on a lattice according to certain rules, the temperature is not a relevant parameter, and the whole dynamics depends on the particle density $\rho$. A dynamical transition may however occur, whose mechanism is quite different from the one which characterizes a thermodynamic phase change; a critical value $\rho_c$ may indeed exist, depending on the model and on the topological properties of the lattice, such that for $\rho>\rho_c$ ergodicity is broken, being the configurational space splitted into two or more irreducible components \cite{TONI}.
\newline
In such a scenario computer simulations are essential to keep track of dynamical aspects, especially when an analytical coherent description of the model is lacking. However, since dynamics is inherently slow, execution times may be really long; here comes the need to speed up simulations by using fast algorithms to provide outputs in suitable times.
\newline
In what follows, we will focus on a widely studied KMC which is the one first proposed by W. Kob and H. C. Andersen \cite{KA}. We will define Markovian dynamics on regular lattices and report the main results obtained so far in an analytical way before describing in detail the underlying logic of the proposed algorithm, that has been used here even to investigate KA dynamics on a more generic topological framework.

\section{The Model}
\label{sec:1}
\subsection{Dynamics on Regular Lattices}
\label{sec:2}
The Kob-Andersen (KA) model is a lattice gas defined on a square lattice $\mathrm{\mathrm{\Lambda}} \in \mathbb{Z}^d$ of size $L^d = N$, where a system of particles interact by hard core repulsion \cite{FMP}. The density $\rho\in\left[0,1\right]$ is a fixed parameter, being the total number of particles conserved by the dynamics, which is time reversible so that detailed balance is satisfied. 
\newline
We indicate with $n_i\left(t\right)\in\left\{0,1\right\}$ the occupation number of site $i$ at time $t$, while $\mathrm{\mathrm{\Omega}}_{N} = \left\{0,1\right\}^{N}$ is the configurational space associated to $\mathrm{\mathrm{\Lambda}}$. Furthermore, we refer to $m\le z-1$ as the activation number of the model, whose role is made clear below, being $z$ the lattice connectivity. 
\newline
Let $\| \cdot \|$ be the usual Euclidean norm. Continuous time Markovian dynamics is explicitly defined as follows:
\begin{itemize}
\item[{-}] choose a particle at random in $\mathrm{\mathrm{\Lambda}}$. Suppose that particle to occupy site $i$; 
\item[{-}] choose with uniform probability $z^{-1}$ one site $j$ among its $z$ neighbors;
\item[{-}] if:
\begin{enumerate}[(a)]
\item $n_j\left(t\right)=0$;
\item $\sum_{\|\mathbf{r}_i - \mathbf{r}_k\|=1} n_k\left(t\right) \le m$;
\item $\sum_{\|\mathbf{r}_j - \mathbf{r}_k\|=1} n_k\left(t\right) \le m+1$;
\end{enumerate}
exchange the occupation numbers of $i$ and $j$, otherwise go back to (a);
\item[{-}] go back to (a).
\end{itemize}
Note that condition (c) guarantees the detailed balance to be satisfied. The parameter $m$ is chosen in a suitable way in order to prevent the particles from diffusing as a free gas with excluded volume ($m=d-1$) and for being completely trapped ($m=0$); usually $m=3$ when $z=6$.
\newline
The backward Markov generator $\mathcal{L}$ acts on local functions $f:\mathrm{\Omega}_N\to \mathbb{R}$ as:
\begin{equation}
\mathcal{L} f\left(\omega\right) =  \sum_{\substack{i,j=1\\ {\|\mathbf{r}_i - \mathbf{r}_j\|=1}}}^N
n_i \left(1-n_j\right) c_{ij}\left(\omega\right) \left[f(\omega^{\left(i,j\right)})-f(\omega)\right]
\end{equation}
where $\omega\in \mathrm{\Omega}_N$ and $\omega^{\left(i,j\right)}$ is the actual configuration $\omega$ in which the occupation numbers of sites $i,j$ are exchanged. The jump rates $c_{ij}\left(\omega\right)$ are defined as:
\begin{equation}
c_{ij}\left(\omega\right) \coloneqq 
\begin{cases}
1 \qquad \text{if (a) and (b) hold true}; \\
0 \qquad \text{otherwise}
\end{cases}
\end{equation}
so that we simply upgrade time of $1/N$ each time a particle moves.
\newline
The same dynamics can be easily extended on different regular graphs like Bethe lattices, which are rooted trees with a fixed number of branches $z$ emanating from each site. We refer to Bethe lattices as infinite volume graphs, while numerical simulations are usually performed on random regular graphs with the same connectivity $z$. These ones are lattices in which each site is randomly connected to other $z=k+1$ sites; no self loops as well as multiple edges are allowed so that they locally look as Bethe lattices, sharing the same statistical property when $N\to \infty$ \cite{WOR}. A classical algorithm, first proposed by B. Bollob\'as \cite{BOLL1,BOLL2}, allows to built up such a graph in a quite simple way. The most relevant difference at finite $N$ regards the possible existence of closed cycles, whose typical length is however of order $\log N$. Dynamical rules are exactly the same as those defined on cubic lattices \cite{TONI}.

\subsection{Characterization of Dynamical Transition}

In the Kob-Andersen model the existence of a thermodynamical phase transition is ruled out by the form of the hamiltonian $\mathcal{H} \coloneqq \sum_{i=1}^N n_i\left(t\right) \equiv N\rho = \text{const}$. The Gibbs measure is flat at any fixed $\rho$, so that all the configurations are equally likely. Moreover, the grand canonical partition function doesn't show any discontinuity, indeed:
\begin{eqnarray}
\mathcal{Z}_{\mu,N,\beta} &=& \sum_{\mathrm{\mathrm{\omega}} \in \mathrm{\mathrm{\Omega}}_{N}} \exp{\left(\beta\left(\mu-1\right)\sum_{i=1}^N n_i^{\left(\mathrm{\mathrm{\omega}}\right)}\right)}
= \sum_{n_1,\ldots,n_N \in \left\{0,1\right\}^N} \exp{\left(\beta\left(\mu-1\right)\sum_{i=1}^N n_i\right)} \nonumber \\
&=& \left[\sum_{k\in\left\{0,1\right\}} \exp\Big(\beta\left(\mu-1\right) k\Big)\right]^N 
= \left[1+\exp\Big(\beta\left(\mu-1\right)\Big)\right]^N.
\end{eqnarray}
A dynamical transition may instead occur when, due to the local constraints, a certain number of configuration in which some particles are forever blocked do exist, so that ergodicity is broken. Such a microscopic mechanism may explain the macroscopic arrest of the dynamic observed in several glassy systems.
\newline
We can keep track of the dynamical transition by looking at some relevant quantities, related to the two and four points correlation functions as defined in Statistical Field Theory \cite{FIELD}, which we expect to diverge as a power law in the thermodynamic limit as $\rho \to \rho_c$. 
\newline
Let 
\begin{equation}
\textsf{q}_i\left(t\right) \coloneqq n_i\left(t\right) n_i\left(0\right)
\end{equation}
and
\begin{equation}
\textsf{q}\left(t\right) \coloneqq \sum_{i=1}^N {\textsf{q}_i\left(t\right)-\rho^2\over  \rho\left(1-\rho\right)}
\end{equation}
be the overlap functions of the model. Define then the density-density correlation function as:
\begin{equation}
\textsf{C}\left(t\right) \coloneqq \left \langle \textsf{q}\left(t\right) \right \rangle 
\end{equation}
where $\left \langle \,\cdot\, \right \rangle$ denotes averages with respect to the whole space of initial configurations $\mathrm{\Omega}_N$. The dynamical susceptibility, which is a measure of the actual size of correlated particles clusters, is defined as:
\begin{equation}
\chi_4 \left(t\right) \coloneqq N \left(\left \langle \textsf{q}^2\left(t\right) \right \rangle - \left \langle \textsf{q}\left(t\right) \right \rangle^2 \right).
\end{equation}
\noindent
On square lattices, it also makes sense to define a space dependent susceptibility \cite{MAR}:
\begin{eqnarray}
\textsf{g}_4\left(r,t\right) &\coloneqq&  
 \sum_{\substack{i,j=1\\ {\|\mathbf{r}_i - \mathbf{r}_j\|=r}}}^N
{\left \langle \textsf{q}_i\left(t\right) \textsf{q}_j\left(t\right) \right \rangle - 
\left \langle \textsf{q}_i\left(t\right) \right \rangle \left \langle \textsf{q}_j\left(t\right) \right \rangle \over N\rho^2 \left(1-\rho^2\right)}.
\end{eqnarray}
This quantity gives informations about decorrelation up to time $t$ between sites at distance $r$. In this position one can express $\chi_4$ as
\begin{equation}
\chi_4 \left(t\right) = \sum_r \textsf{g}_4\left(r,t\right).
\end{equation}
Note that in the ergodic phase, due to the time decorrelation among $n_i\left(0\right)$ and $n_i\left(t\right)$ at $t\gg 1$ \cite{RS}:
\begin{eqnarray}
\lim_{N\to \infty} \lim_{t\to \infty} &\textsf{C}\left(t\right)& = 0 \label{ci} \\ 
\lim_{N\to \infty} \lim_{t\to \infty} &\chi_4\left(t\right)& = 1 \label{chi}.
\end{eqnarray}

\subsection{A Few Analytic Results}

We here list some results concerning the KA model defined on square lattices and Bethe lattices at fixed connectivity $d$ when $N\to \infty$. 
\newline
The finite volume case is trivial in the sense that we can always find a critical density $\rho_c\left(N\right)$, which depends on the size of the system, such that for $\rho>\rho_c\left(N\right)$ the Markovian dynamics turns out to be reducible \cite{TONI,BERT}.
\newline
However, it can be shown that on square lattices $\mathbb{Z}^d$ ergodicity is an inner property of the system for any $\rho\in\left[0,1\right)$ provided the activation parameter to be such that $m>d-1$ \cite{TONI}.
\newline
Regarding the Bethe dynamics, a critical point $\rho_c$ does actually exist, whose value, which depends on $k$ and $m$, can be found by solving a system of coupled polynomial equations \cite{TONI}. In the case in which parameters are chosen in order to investigate any similarities with the model defined on a square lattice in $d=3$, i.e. when $k=5$, one gets $\rho_c \simeq 0.835$.
\newline
Numerical simulations clearly take place on finite volume lattices. We partially get rid of finite size effects by adopting periodic boundary condition in the case of cubic lattices and by moving the whole dynamics on a random regular graph in the case of Bethe lattices as mentioned above. Although this convention works good in the latter case, it fails in eliminating tracks of the transition which occur when $N<\infty$ in the first one. Moreover, $\rho_c\left(N\right)$ depends weakly on $N$ being $\rho_c\left(N\right) = 1-C/\log\left(-d\log N\right)$, where $C$ is a suitable constant \cite{TONI}, so that its value is not very susceptible to a volume increase. For $N=\mathcal{O}\left(10^5\right)$, $\rho_c\left(N\right) \simeq 0.881$ \cite{KA}.

\section{The Algorithm}

\subsection{Multispin Coding}

In what follows we implicitly refer to the KA model on graphs with connectivity $z=6$ being fixed the activation number to $m=3$.
\newline
Multispin techniques provide a natural way to speed up numerical simulations on binary lattices. The main idea, which is pretty simple, consists in suitably exploiting the available memory of a given machine by coding in a boolean logic. 
\newline
The most natural way in order to write a working algorithm for the KA dynamics may follow classical steps, at $\rho$ fixed. For example:
\begin{itemize}
\item select $\omega \in \mathrm{\mathrm{\Omega}}_{N,\rho}$ at random;
\item run the KA dynamics and compute correlations of type $\textsf{q}_i\left(t\right)\textsf{q}_j\left(t\right)$;
\item do the same for certain number of initial configurations;
\item compute averages.
\end{itemize}
Roughly speaking, a multispin approach enables to run dynamics in parallel for a number of initial configuration equal to the number of bits corresponding to the specific programming language type chosen to store at each step the actual value of the occupation number. More specifically, suppose to work with types which reserve $N_{b}$ bits in memory. Since $n_i\left(t\right)\in\left\{0,1\right\}$ we could think to build an ordered sequence in such a way that each bit carries information about the occupation number of the $i$-th site of $N_b$ different lattices. By using bit-a-bit operators in order to manipulate these object, we should expect to reduce the execution time by a factor of order $N_b$\footnote{This is approximately true when $N_b$ is not large too much. We have indeed to take into account the computational cost due to specific logical functions whose execution time increases with $N_b$.}.

\begin{figure}[!htbp]
\centering
\begin{adjustbox}{max width=0.65\textwidth}
\begin{bytefield}{32}
\bitheader[endianness=big]{0-31} \\
\bitbox{1}{0} & \bitbox{1}{0} & \bitbox{1}{0} & \bitbox{1}{0} &
\bitbox{1}{0} & \bitbox{1}{0} & \bitbox{1}{0} & \bitbox{1}{0} &
\bitbox{1}{0} & \bitbox{1}{0} & \bitbox{1}{0} & \bitbox{1}{0} &
\bitbox{1}{0} & \bitbox{1}{0} & \bitbox{1}{0} & \bitbox{1}{0} &
\bitbox{1}{0} & \bitbox{1}{0} & \bitbox{1}{0} & \bitbox{1}{0} &
\bitbox{1}{0} & \bitbox{1}{0} & \bitbox{1}{0} & \bitbox{1}{0} &
\bitbox{1}{0} & \bitbox{1}{0} & \bitbox{1}{0} & \bitbox{1}{0} &
\bitbox{1}{0} & \bitbox{1}{0} & \bitbox{1}{0} & \bitbox{1}{1} 
\end{bytefield}
\end{adjustbox}
\end{figure}
\vspace{-0.5cm}
\begin{figure}[!htbp]
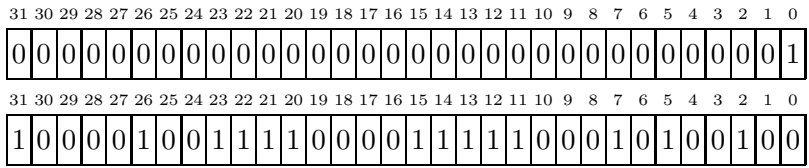

\centering
\begin{adjustbox}{max width=0.65\textwidth}
\begin{bytefield}{32}
\bitheader[endianness=big]{0-31} \\
\bitbox{1}{1} & \bitbox{1}{0} & \bitbox{1}{0} & \bitbox{1}{0} &
\bitbox{1}{0} & \bitbox{1}{1} & \bitbox{1}{0} & \bitbox{1}{0} &
\bitbox{1}{1} & \bitbox{1}{1} & \bitbox{1}{1} & \bitbox{1}{1} &
\bitbox{1}{0} & \bitbox{1}{0} & \bitbox{1}{0} & \bitbox{1}{0} &
\bitbox{1}{1} & \bitbox{1}{1} & \bitbox{1}{1} & \bitbox{1}{1} &
\bitbox{1}{1} & \bitbox{1}{0} & \bitbox{1}{0} & \bitbox{1}{0} &
\bitbox{1}{1} & \bitbox{1}{0} & \bitbox{1}{1} & \bitbox{1}{0} &
\bitbox{1}{0} & \bitbox{1}{1} & \bitbox{1}{0} & \bitbox{1}{0} 
\end{bytefield}
\end{adjustbox}
\caption{Schematic representation of a 32-bits type. The first row carries information of just one occupation number of a certain site. In the second row instead, 32 occupation numbers corresponding to 32 different initial conditions are stored.}
\end{figure}

\subsection{Updating}

Despite the simplicity of the multispin technique idea, the coding process may be a non trivial issue as well as the computation of the observables. Indeed, at each step we have to update the occupation numbers according to the dynamical rules as defined in Section 1 by using logical operators. 
\newline
By denoting from now on with $\texttt{n}_i$ the whole string of the $N_b$ occupation numbers of the $i$-th site, and with $\texttt{n}_i^{\left(b\right)}$ the actual value of the corresponding $b$-th lattice, we can resume as follows the updating logic.
\newline
Choose random sites $i$ and $j$, being $i$, $j$ nearest neighbors, and define strings $\texttt{s}_{i_4}$, $\texttt{s}_{j_4}$ and $\texttt{s}_{i_{5}}$, $\texttt{s}_{j_{5}}$ according to the convention:
\begin{enumerate}
\item[{-}] $\texttt{s}_{k_4}^{\left(b\right)}=1$ iff the $k$-the site of the $b$-th lattice has a number of occupied neighbors less or equal to $m=3$;
\item[{-}] $\texttt{s}_{k_{5}}^{\left(b\right)}=1$ iff the $k$-the site of the $b$-th lattice has a number of occupied neighbors less or equal to $m=4$.
\end{enumerate} 
Such sequences may be constructed starting from Karnaugh maps \cite{KAR} by using bit-a-bit operators as follows:
\begin{eqnarray}
\texttt{s}_{k_4} =
\neg &\bigvee_{\substack{z_{\alpha},z_{\beta},z_{\gamma},z_{\delta}\in \left\{\text{n.n.}\,k\right\}\\ {z_{\alpha} \neq z_{\beta} \neq z_{\gamma} \neq z_{\delta}}}}&
\texttt{n}_{z_{\alpha}}\wedge\texttt{n}_{z_{\beta}}\wedge\texttt{n}_{z_{\gamma}}\wedge\texttt{n}_{z_{\delta}}  \nonumber \\ \\ \nonumber \\
\texttt{s}_{k_5} = 
\neg &\bigvee_{\substack{z_{\alpha},z_{\beta},z_{\gamma},z_{\delta},z_{\varepsilon}\in \left\{\text{n.n.}\,k\right\}\\ {z_{\alpha} \neq z_{\beta} \neq z_{\gamma} \neq z_{\delta} \neq z_{\varepsilon}}}}&
\texttt{n}_{z_{\alpha}}\wedge\texttt{n}_{z_{\beta}}\wedge\texttt{n}_{z_{\gamma}}\wedge\texttt{n}_{z_{\delta}} \wedge \texttt{n}_{z_{\varepsilon}}. \nonumber \\
\end{eqnarray}
Compute then:
\begin{eqnarray}
\texttt{s}_i &=& \texttt{s}_{i_4} \wedge \texttt{s}_{j_5} \\
\texttt{s}_j &=& \texttt{s}_{j_4} \wedge \texttt{s}_{i_5}
\end{eqnarray}
and update $(\texttt{n}_i,\texttt{n}_j) \mapsto (\texttt{n}_i^+,\texttt{n}_j^+)$ according to table \ref{tabUp}. Explicitly:
\begin{eqnarray}
\texttt{n}_i^+ &=& \left(\texttt{n}_i \wedge \texttt{n}_j \right) \vee \left(\texttt{n}_j \wedge \texttt{s}_j\right)
\vee \left(\texttt{n}_i \wedge \neg \texttt{s}_i\right) \\
\texttt{n}_j^+ &=& \left(\texttt{n}_i \wedge \texttt{n}_j \right) \vee \left(\texttt{n}_i \wedge \texttt{s}_i\right)
\vee \left(\texttt{n}_j \wedge \neg \texttt{s}_j\right).
\end{eqnarray}
The operations above complete a single dynamical step.

\begin{table}[!htbp]
\centering
\begin{tabular}{cccc|c|c}
$\texttt{n}_i$ & $\texttt{n}_j$ & $\texttt{s}_i $ & $\texttt{s}_j$ & $\texttt{n}_i^+$ & $\texttt{n}_j^+$  \\
\hline
0 & 0 & 0 & 0       & 0 & 0\\
\hline
0 & 0 & 0 & 1       & 0 & 0\\
\hline
0 & 0 & 1 & 0       & 0 & 0\\
\hline
0 & 0 & 1 & 1       & 0 & 0\\
\hline
0 & 1 & 0 & 0       & 0 & 1\\
\hline
0 & 1 & 0 & 1       & 1 & 0\\
\hline
0 & 1 & 1 & 0       & 0 & 1\\
\hline
0 & 1 & 1 & 1       & 1 & 0\\
\hline
1 & 0 & 0 & 0       & 1 & 0\\
\hline
1 & 0 & 0 & 1       & 1 & 0\\
\hline
1 & 0 & 1 & 0       & 0 & 1\\
\hline
1 & 0 & 1 & 1       & 0 & 1\\
\hline
1 & 1 & 0 & 0       & 1 & 1\\
\hline
1 & 1 & 0 & 1       & 1 & 1\\
\hline
1 & 1 & 1 & 0       & 1 & 1\\
\hline
1 & 1 & 1 & 1       & 1 & 1
\end{tabular}
\caption{Truth table corresponding to the updating $(\texttt{n}_i,\texttt{n}_j) \mapsto (\texttt{n}_i^+,\texttt{n}_j^+)$.}\label{tabUp}
\end{table}

\subsection{Observables}

We here suppose to perform averages on exactly $N_b$ initial configurations. Extension to multiples of $N_b$ is technically straightforward. 
\newline
Let be:
\begin{equation}
\varphi_{ij}\left(t\right) \coloneqq {\mathbf{1}}_{\left\{n_i\left(t\right)n_i\left(0\right)n_j\left(t\right)n_j\left(0\right) = 1\right\}}
\end{equation}
which can be numerically computed as:
\begin{equation}\label{corr}
\varphi_{ij}\left(t\right) = \texttt{n}_i\texttt{[t]} \wedge \texttt{n}_i\texttt{[0]} \wedge \texttt{n}_j\texttt{[t]} \wedge
\texttt{n}_j\texttt{[0]}
\end{equation}
and call $\overline{\varphi}_{ij}\left(t\right)$ the same quantity averaged on the $N_b$ different initial configurations. Notice that $\overline{\varphi}_{ij}\left(t\right)$ actually corresponds to the sum of the ones contained in the string \eqref{corr}. Define:
\begin{equation}
\mathrm{\Phi}\left(r,t\right) \coloneqq  \sum_{\substack{i,j=1\\ {\|\mathbf{r}_i - \mathbf{r}_j\|=r}}}^N \overline{\varphi}_{ij}\left(t\right)
\end{equation}
where, with a little abuse of notation, $\| \cdot \|$ stands for the Euclidean norm on cubic lattices while 
on a tree-like graph ${\|\mathbf{r}_i - \mathbf{r}_j\|=0}$ if $j=i$ and ${\|\mathbf{r}_i - \mathbf{r}_j\|=1}$ otherwise. According to this convention we get:
\begin{eqnarray}
\textsf{C}\left(t\right) &=&  {\mathrm{\mathrm{\Phi}}\left(0,t\right) - N\rho^2 \over N\rho \left(1-\rho\right)} \\
\chi_4\left(t\right) &=& {\sum_{r > 0}2\mathrm{\Phi}\left(r,t\right) + \mathrm{\Phi}\left(0,t\right) - \mathrm{\Phi}^2\left(0,t\right) 
\over N\rho^2 \left(1-\rho\right)^2} \\
\textsf{g}_4\left(r,t\right) &=& {\mathrm{\Phi}\left(r,t\right) - {z\over 2N}\mathrm{\Phi}^2\left(0,t\right)\over N\rho^2 \left(1-\rho\right)^2}.
\end{eqnarray}

\section{Numerical Results}

\subsection{Cubic Lattices}

We tested the algorithm by running simulations on a square lattice of side $L=20$ in $d=3$ for several values of $\rho$ in the ergodic phase. We expect the dynamics to slow down as $\rho$ increases until $\rho_c\left(N\right)$. At $\rho>\rho_c\left(N\right)$ clusters of blocked particles do appear and ergodicity is broken \cite{KA,TONI,FMP,MAR}.


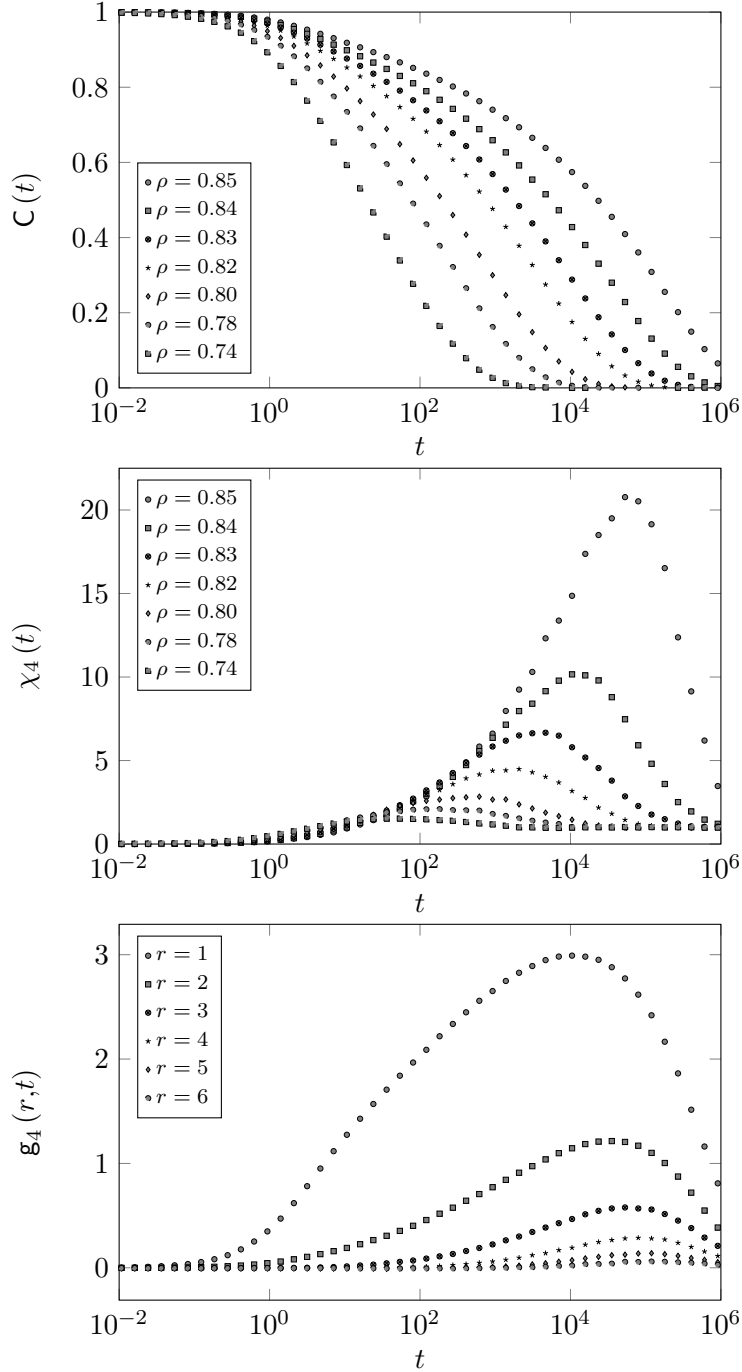
\begin{figure}[!htbp]
\centering
\begin{tikzpicture}
\begin{axis}[
width=8cm,
height=5cm,
scale only axis,
ylabel=$\textsf{C}\left(t\right)$,
xlabel=$t$,
xmin=0.01,
xmax=1000000,
xmode=log,
log basis x={10},
xtick={0.0001,0.01,1,100,10000,1000000},
ymin=0,
ymax=1,
legend style={draw=black,thin,at={(0.03,0.6)},anchor=north west,legend cell align=left,font=\scriptsize},
]
\addplot+[only marks,mark size=0.35mm,mark options={line width=0.025pt,fill=gray},color=black]
table{chi2cubic85.dat};
\addlegendentry{$\rho=0.85$};
\addplot+[only marks,mark size=0.35mm,mark options={line width=0.025pt,fill=gray},color=black]
table{chi2cubic84.dat};
\addlegendentry{$\rho=0.84$};
\addplot+[only marks,mark size=0.35mm,mark options={line width=0.025pt,fill=gray},color=black]
table{chi2cubic83.dat};
\addlegendentry{$\rho=0.83$};
\addplot+[only marks,mark size=0.35mm,mark options={line width=0.025pt,fill=gray},color=black]
table{chi2cubic82.dat};
\addlegendentry{$\rho=0.82$};
\addplot+[only marks,mark size=0.35mm,mark options={line width=0.025pt,fill=gray},color=black]
table{chi2cubic80.dat};
\addlegendentry{$\rho=0.80$};
\addplot+[only marks,mark size=0.35mm,mark options={line width=0.025pt,fill=gray},color=black]
table{chi2cubic78.dat};
\addlegendentry{$\rho=0.78$};
\addplot+[only marks,mark size=0.35mm,mark options={line width=0.025pt,fill=gray},color=black]
table{chi2cubic74.dat};
\addlegendentry{$\rho=0.74$};
\end{axis}
\end{tikzpicture}

\begin{tikzpicture}
\begin{axis}[
width=8cm,
height=5cm,
scale only axis,
ylabel=$\chi_4\left(t\right)$,
xlabel=$t$,
xmin=0.01,
xmax=1000000,
xmode=log,
log basis x={10},
xtick={0.0001,0.01,1,100,10000,1000000},
ymin=0,
ymax=22.5,
ytick={0,5,10,15,20},
legend style={draw=black,thin,at={(0.03,0.97)},anchor=north west,legend cell align=left,font=\scriptsize},
]
\addplot+[only marks,mark size=0.35mm,mark options={line width=0.025pt,fill=gray},color=black]
table{chi4cubic85.dat};
\addlegendentry{$\rho=0.85$};
\addplot+[only marks,mark size=0.35mm,mark options={line width=0.025pt,fill=gray},color=black]
table{chi4cubic84.dat};
\addlegendentry{$\rho=0.84$};
\addplot+[only marks,mark size=0.35mm,mark options={line width=0.025pt,fill=gray},color=black]
table{chi4cubic83.dat};
\addlegendentry{$\rho=0.83$};
\addplot+[only marks,mark size=0.35mm,mark options={line width=0.025pt,fill=gray},color=black]
table{chi4cubic82.dat};
\addlegendentry{$\rho=0.82$};
\addplot+[only marks,mark size=0.35mm,mark options={line width=0.025pt,fill=gray},color=black]
table{chi4cubic80.dat};
\addlegendentry{$\rho=0.80$};
\addplot+[only marks,mark size=0.35mm,mark options={line width=0.025pt,fill=gray},color=black]
table{chi4cubic78.dat};
\addlegendentry{$\rho=0.78$};
\addplot+[only marks,mark size=0.35mm,mark options={line width=0.025pt,fill=gray},color=black]
table{chi4cubic74.dat};
\addlegendentry{$\rho=0.74$};
\end{axis}
\end{tikzpicture}

\begin{tikzpicture}
\begin{axis}[
width=8cm,
height=5cm,
scale only axis,
ylabel=$\textsf{g}_4\left(r \text{,} t \right)$,
xlabel=$t$,
xmin=0.01,
xmax=1000000,
xmode=log,
log basis x={10},
xtick={0.0001,0.01,1,100,10000,1000000},
legend style={draw=black,thin,at={(0.03,0.97)},anchor=north west,legend cell align=left,font=\scriptsize},
]
\addplot+[only marks,mark size=0.35mm,mark options={line width=0.025pt,fill=gray},color=black]
table{pippo1.dat};
\addlegendentry{$r=1$};
\addplot+[only marks,mark size=0.35mm,mark options={line width=0.025pt,fill=gray},color=black]
table{pippo2.dat};
\addlegendentry{$r=2$};
\addplot+[only marks,mark size=0.35mm,mark options={line width=0.025pt,fill=gray},color=black]
table{pippo3.dat};
\addlegendentry{$r=3$};
\addplot+[only marks,mark size=0.35mm,mark options={line width=0.025pt,fill=gray},color=black]
table{pippo4.dat};
\addlegendentry{$r=4$};
\addplot+[only marks,mark size=0.35mm,mark options={line width=0.025pt,fill=gray},color=black]
table{pippo5.dat};
\addlegendentry{$r=5$};
\addplot+[only marks,mark size=0.35mm,mark options={line width=0.025pt,fill=gray},color=black]
table{pippo6.dat};
\addlegendentry{$r=6$};
\end{axis}
\end{tikzpicture}
\caption{Density-density correlation function and dynamical susceptibility for $\rho$ ranging from $0.74$ to $0.85$. Typical times increases with $\rho$ as well as the maximum of $\chi_4$. At the bottom, $\textsf{g}_4\left(r,t\right)$ at fixed $\rho=0.85$.}\label{obsKA}
\end{figure}


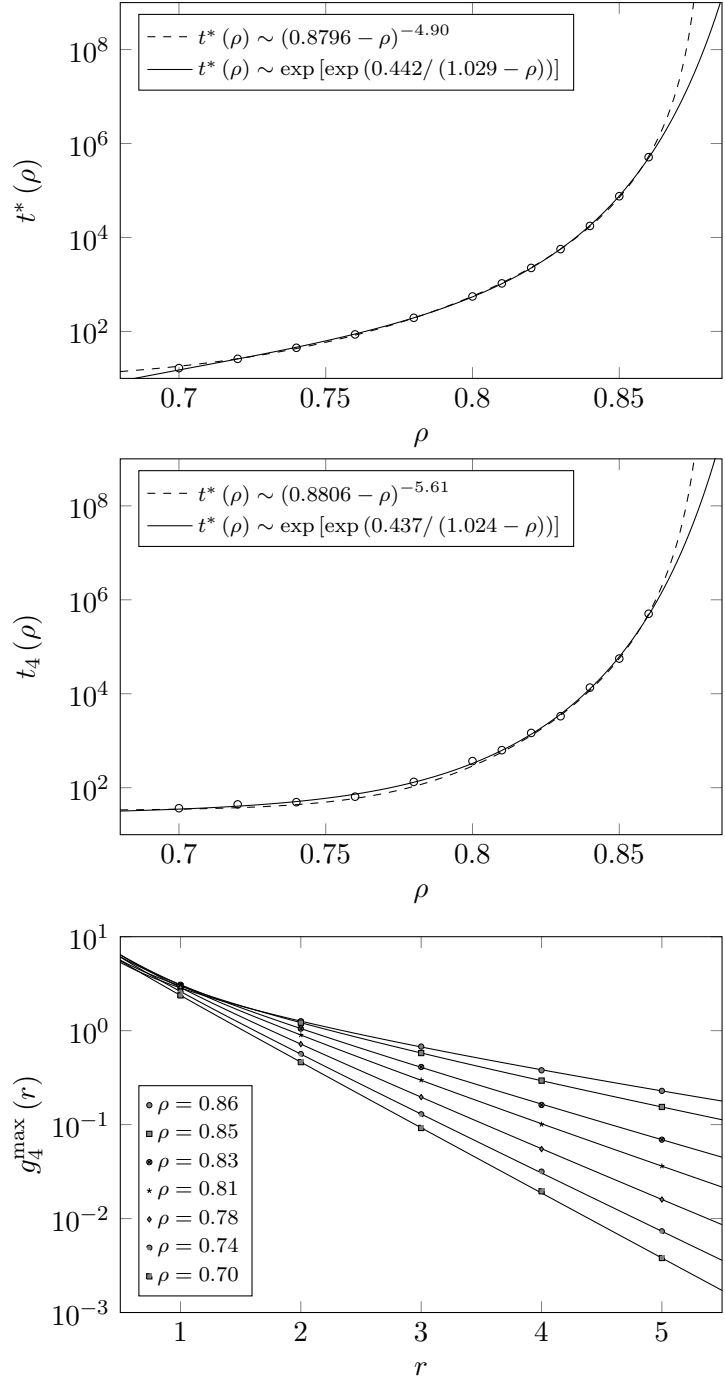
\begin{figure}[htbp!]
\centering
\begin{tikzpicture}
\begin{axis}[
width=8cm,
height=5cm,
scale only axis,
ylabel=$t^*\left(\rho\right)$,
xlabel=$\rho$,
ymin=10,
ymax=1000000000,
xmin=0.68,
xmax=0.885,
ymode=log,
log basis y={10},
xtick={0.70,0.75,0.80,0.85},
ytick={100,10000,1000000,100000000},
legend style={draw=black,thin,at={(0.03,0.97)},anchor=north west,legend cell align=left,font=\scriptsize},
]
\addplot[dashed,domain=0.68:0.885,samples=200]
{7.88401+0.00226649*(1/(0.879591-x))^4.90319};
\addlegendentry{$t^*\left(\rho\right) \sim \left(0.8796-\rho\right)^{-4.90} $};
\addplot[domain=0.68:0.885,samples=200]
{-10.8466 + 0.561479*exp(exp(0.442147/(1.02899-x)))};
\addlegendentry{$t^*\left(\rho\right) \sim \exp \left[\exp\left({0.442/\left(1.029 -\rho\right)}  \right)\right]$};
\addplot[only marks,mark=o,mark size=0.50mm,color=black]
table{tauKA.dat};
\end{axis}
\end{tikzpicture}

\begin{tikzpicture}
\begin{axis}[
width=8cm,
height=5cm,
scale only axis,
ylabel=$t_4\left(\rho\right)$,
xlabel=$\rho$,
ymin=10,
ymax=1000000000,
xmin=0.68,
xmax=0.885,
ymode=log,
log basis y={10},
xtick={0.70,0.75,0.80,0.85},
ytick={100,10000,1000000,100000000},
legend style={draw=black,thin,at={(0.03,0.97)},anchor=north west,legend cell align=left,font=\scriptsize},
]
\addplot[dashed,domain=0.68:0.885,samples=200]
{32.1161+0.000191616*(1/(0.880972-x))^5.61267};
\addlegendentry{$t^*\left(\rho\right) \sim \left(0.8806-\rho\right)^{-5.61} $};
\addplot[domain=0.68:0.885,samples=200]
{22.3634 + 0.267689*exp(exp(0.437257/(1.02401-x)))};
\addlegendentry{$t^*\left(\rho\right) \sim \exp \left[\exp\left({0.437/\left(1.024 -\rho\right)}  \right)\right]$};
\addplot[only marks,mark=o,mark size=0.50mm,color=black]
table{t_KA.dat};
\end{axis}
\end{tikzpicture}

\begin{tikzpicture}
\begin{axis}[
width=8cm,
height=5cm,
scale only axis,
ylabel=$g_4^{\text{max}}\left(r\right)$,
xlabel=$r$,
xmin=0.5,
xmax=5.5,
ymode=log,
log basis y={10},
xtick={1,2,3,4,5},
ytick={0.001,0.01,0.1,1,10},
ymin=0.001,
ymax=10,
legend style={draw=black,thin,at={(0.03,0.605)},anchor=north west,legend cell align=left,font=\scriptsize},
]
\addplot+[only marks, mark size=0.35mm,mark options={line width=0.025pt,fill=gray},color=black]
table{g86.dat};
\addlegendentry{$\rho=0.86$};
\addplot+[only marks,mark size=0.35mm,mark options={line width=0.025pt,fill=gray},color=black]
table{g85.dat};
\addlegendentry{$\rho=0.85$};
\addplot+[only marks,mark size=0.35mm,mark options={line width=0.025pt,fill=gray},color=black]
table{g83.dat};
\addlegendentry{$\rho=0.83$};
\addplot+[only marks,mark size=0.35mm,mark options={line width=0.025pt,fill=gray},color=black]
table{g81.dat};
\addlegendentry{$\rho=0.81$};
\addplot+[only marks,mark size=0.35mm,mark options={line width=0.025pt,fill=gray},color=black]
table{g78.dat};
\addlegendentry{$\rho=0.78$};
\addplot+[only marks,mark size=0.35mm,mark options={line width=0.025pt,fill=gray},color=black]
table{g74.dat};
\addlegendentry{$\rho=0.74$};
\addplot+[only marks,mark size=0.35mm,mark options={line width=0.025pt,fill=gray},color=black]
table{g70.dat};
\addlegendentry{$\rho=0.70$};
\addplot[thin,domain=0.5:5.5,samples=200,color=black]
{(4.15805/(x^0.631764))*exp(-x/2.65307)};
\addplot[thin,domain=0.5:5.5,samples=200,color=black]
{(5.03600/(x^0.539591))*exp(-x/1.90972)};
\addplot[thin,domain=0.5:5.5,samples=200,color=black]
{(6.70616/(x^0.428126))*exp(-x/1.28674)};
\addplot[thin,domain=0.5:5.5,samples=200,color=black]
{(7.90411/(x^0.390652))*exp(-x/1.0508)};
\addplot[thin,domain=0.5:5.5,samples=200,color=black]
{(9.33198/(x^0.305567))*exp(-x/0.850545)};
\addplot[thin,domain=0.5:5.5,samples=200,color=black]
{(10.3595/(x^0.22411))*exp(-x/0.725226)};
\addplot[thin,domain=0.5:5.5,samples=200,color=black]
{(11.4297/(x^0.0907917))*exp(-x/0.635979)};
\end{axis}
\end{tikzpicture}
\caption{(First two figures) Characteristic times as functions of $\rho$. A double exponential fit seems to work well. (Last figure) Maximum of $\textsf{g}_4\left(r,t\right)$ for several values of $\rho$. For each $\rho$, the solid lines correspond to exponential fits according to \eqref{fitg}.}\label{gmax}
\end{figure}

\newpage
\noindent
From the top to the bottom, in Figure \ref{obsKA} we show the behavior of the density-density correlation function, the dynamical susceptibility and the four point spatial correlated correlation function at fixed $\rho=0.85$. Notice in particular that the maximum of $\chi_4$ increases with $\rho$. We extracted from $\textsf{C}\left(t\right)$ and $\chi_4\left(t\right)$ decorrelation times $t^*\left(\rho\right)$, $t_4\left(\rho\right)$ respectively defined by $\textsf{C}\left(t^*\right) = \exp\left(-1\right)$, $t_4\left(\rho\right) \coloneqq \underset{t}{\arg\max} \, \chi_4\left(t\right)$.
\newline
These are quantities we expect to diverge as $\rho$ approaches $\rho_c\left(N\right)\simeq 0.881$. This is actually the case if we look at Figure 3, where $t^*\left(\rho\right) - \rho$ and $t_4\left(\rho\right) - \rho$ trends are shown. 
\noindent
A correlation length $\xi_4\left(\rho\right)$ can be even extracted from $\textsf{g}_4\left(r,t\right)$ by supposing the decay to be exponential:
\begin{equation}\label{fitg}
\textsf{g}_4^{\text{max}}\left(r\right) \sim r^{-\lambda\left(\rho\right)} \exp\left(-{r\over \xi_4\left(\rho\right)}\right).
\end{equation}
We fitted data according to \eqref{fitg} (Figure 4). For $\rho$ ranging in $\left[0.70,0.86\right]$ we obtained values for $\lambda\left(\rho\right)$ and $\xi_4\left(\rho\right)$ respectively ranging in $\left[0.0908,0.632\right]$ and $\left[0.636,2.65\right]$ which are in good agreement with those reported in \cite{MAR}.

\subsection{Dynamics on Random Regular Graphs}

On Bethe lattices a dynamical transition does actually exist, so that we expect decorrelation times $t^*\left(\rho\right)$ and $t_4\left(\rho\right)$ to diverge as a true power law at $\rho_c$:
\begin{eqnarray}
t^*\left(\rho\right) &\sim& \left(\rho_c - \rho\right)^{-\eta} \\
t_4\left(\rho\right) &\sim& \left(\rho_c - \rho\right)^{-\nu}.
\end{eqnarray}
Following literally \cite{TONI}, consider a site $i$ and define the following events:
\begin{enumerate}
\item[(i)] the site is occupied by a particle which can never move up as long as the site below is occupied; denote by $Y\left(\rho\right)$ the probability of this event;
\newline
\item[(ii)] the site is frozen: occupied by a particle which can never move up even if the site below is empty; denote by $F\left(\rho\right)$ the probability of this event, which is a subset of (i);
\newline
\item[(iii)] the site is empty but blocked in such a way that particles on lower sites can never move up to this site; denote by $B\left(\rho\right)$ the probability of this event.
\end{enumerate}
By using the tree-like structure and its symmetry under exchange of branches, one can write iterative equations for the (primed) probabilities for the site in one layer, in terms of the (unprimed) probabilities for the sites in the layer immediately above, which in the case $k=5$, $m=3$ are:
\begin{empheq}[left=\empheqlbrace]{align}\label{snewton}
Y' &= F + 10\rho Y^3 \left[\left(1-Y\right)^2 - B^2\right] \nonumber    \\ 
B' &= \left(1-\rho\right) F^4 \left(5-4F \right)   \\ 
F' &= \rho \left[\left(Y+B\right)^5 + 5Y^4 \left(1-Y-B\right)\right]. \nonumber
\end{empheq}
We refer to $\rho_c$ as the value for which a couple of fixed point does appear beside the trivial solution $\left(Y,B,F\right)\equiv\left(0,0,0\right)$ (see \cite{TONI} for further details). 
\newline
When solving numerically \eqref{snewton}, we obtain (Figure \ref{grande}):
\begin{empheq}[left=\empheqlbrace]{align}
\rho_c &\simeq 0.835 \nonumber \\ 
Y_c &\simeq 0.725 \nonumber \\ 
B_c &\simeq 0.0283 \nonumber \\
F_c &\simeq 0.487. \nonumber
\end{empheq}


\begin{figure}[htbp!]
\centering
\begin{tikzpicture}[trim axis left, trim axis right]
\begin{axis}[
width=8cm,
height=10cm,
scale only axis,
xlabel=$\rho$,
ymin=0,
ymax=1,
xmin=0.8,
xmax=1,
xtick={0.8,0.85,0.9,0.95,1},
ytick={0.2,0.4,0.6,0.8,1},
legend style={draw=black,thin,at={(0.03,0.975)},anchor=north west,font=\scriptsize},
]
\addplot[solid,no marks,very thin,color=black]
table{ybisBif.dat};
\addlegendentry{$Y\left(\rho\right)$};
\addplot[dashed,no marks,color=black]
table{bBif.dat};
\addlegendentry{$B\left(\rho\right)$};
\addplot[dotted,no marks,thick,color=black]
table{fBif.dat};
\addlegendentry{$F\left(\rho\right)$};
\addplot[only marks,mark=*,mark size=0.30mm,color=black] coordinates {(0.83477,0.724917)};
\addplot[only marks,mark=*,mark size=0.30mm,color=black] coordinates {(0.83477,0.0283342)};
\addplot[only marks,mark=*,mark size=0.30mm,color=black] coordinates {(0.83477,0.486831)};
\draw[dashed,thin,color=gray] (axis cs:0.83477,-0.05) -- (axis cs:0.83477,0.724917);
\draw[dashed,thin,color=gray] (axis cs:0,0.724917) -- (axis cs:0.83477,0.724917);
\draw[dashed,thin,color=gray] (axis cs:0,0.0283342) -- (axis cs:0.83477,0.0283342);
\draw[dashed,thin,color=gray] (axis cs:0,0.486831) -- (axis cs:0.83477,0.486831);
\end{axis}
\begin{axis}[
axis lines=none,
width=8cm,
height=10cm,
scale only axis,
ymin=0,
ymax=1,
xmin=0.8,
xmax=1,
xtick={1.5},
ytick={1.5},
clip=false,]
\node at (axis cs:0.793,0.724917) {$Y_c$};
\node at (axis cs:0.793,0.0283342) {$B_c$};
\node at (axis cs:0.793,0.486831) {$F_c$};
\node at (axis cs:0.83477,-0.025) {$\rho_c$};
\end{axis}
\end{tikzpicture}
\caption{Bifurcations of the map \eqref{snewton}. At $\rho>\rho_c$ a stable solution does appear beside the trivial one (upper branch) together with an unstable one (bottom).}\label{grande}
\end{figure}
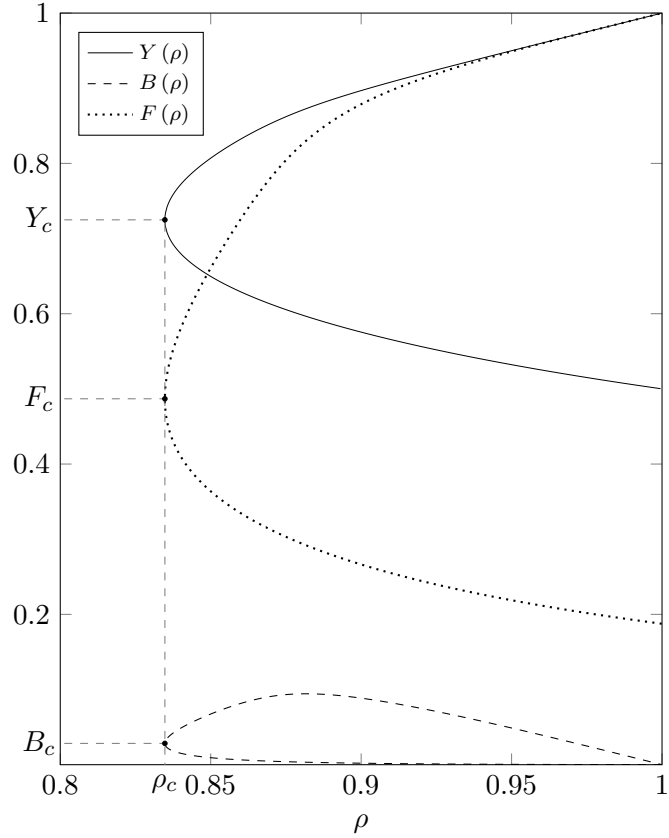

\noindent
Data seem to be well fitted by a double exponential function with a singularity close to $1$ rather than a power law with an empirical apparent critical density at $\rho\simeq 0.881$. This fact can be interpreted as a sign of the lacking of a dynamical transition as $N\to \infty$. The KA dynamics has then been run on a random regular graph with $N=10^5$ and connectivity $z=6$ extracted uniformly at random among the lattices of the same class. As in the case of square lattices, the measured $\textsf{C}\left(t\right)$ and $\chi_4\left(t\right)$ exhibit a typical trend. We then extrapolated correlation times for several values of $\rho<\rho_c\simeq 0.835$ in order to obtain empirical estimates for the critical point (Figure \ref{corrBL}), which are in good agreement with the expected theoretical value.


\begin{figure}[t]
\centering
\begin{tikzpicture}
\begin{axis}[
width=8cm,
height=5cm,
scale only axis,
ylabel=$t^*\left(\rho\right)$,
xlabel=$\rho$,
ymin=10,
ymax=1000000000,
xmin=0.68,
xmax=0.83477,
ymode=log,
log basis y={10},
ytick={100,10000,1000000,100000000},
legend style={draw=black,thin,at={(0.03,0.97)},anchor=north west,legend cell align=left,font=\scriptsize},
]
\addplot[dashed,domain=0.68:0.83477,samples=200]
{1.16674+0.0438459*(1/(0.834326-x))^2.81828};
\addlegendentry{$t^*\left(\rho\right) \sim \left(0.8343-\rho\right)^{-2.82} $};
\addplot[only marks,mark=o,mark size=0.50mm,mark options={line width=0.025pt,fill=white},color=black]
table{tauBL.dat};
\end{axis}
\end{tikzpicture}
\begin{tikzpicture}
\begin{axis}[
width=8cm,
height=5cm,
scale only axis,
ylabel=$t_4\left(\rho\right)$,
xlabel=$\rho$,
ymin=10,
ymax=1000000000,
xmin=0.68,
xmax=0.83477,
ymode=log,
log basis y={10},
ytick={100,10000,1000000,100000000},
legend style={draw=black,thin,at={(0.03,0.97)},anchor=north west,legend cell align=left,font=\scriptsize},
]
\addplot[dashed,domain=0.68:0.83477,samples=200]
{10.8562+0.0139507*(1/(0.833474-x))^3.23545};
\addlegendentry{$t_4\left(\rho\right) \sim \left(0.8334-\rho\right)^{-3.23} $};
\addplot[only marks,mark=o,mark size=0.50mm,mark options={line width=0.025pt,fill=white},color=black]
table{tBL.dat};
\end{axis}
\end{tikzpicture}
\caption{Behavior of $t^*\left(\rho\right)$ and $t_4\left(\rho\right)$ at varying $\rho<\rho_c$.}\label{corrBL}
\end{figure}
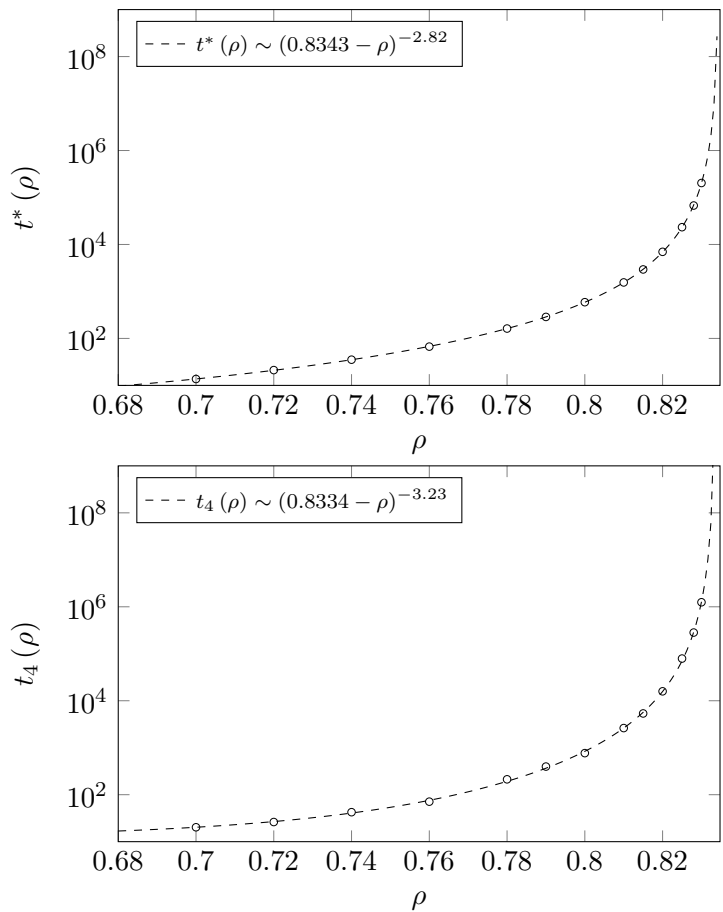

\section{An Interpolating Model}

\subsection{The Conjecture}

The existence of an ergodic/non-ergodic transition in the thermodynamic limit seems to be due to certain topological properties which characterize the graph on which KA dynamics takes place, being the connectivity as well as geometrical constraints the same in the two cases analyzed so far. In particular, we conjecture this to be connected to the average number of closed paths of the reference graph as well as their average minimum length, which is of order 1 in the case of a square lattice while it is zero for a Bethe lattice. Indeed, due to the dynamical rules, the particles moving on a tree graph may be more easily trapped in a frozen configuration since they have fewer escape routes.
\newline
We are thus looking for a graph in which these geometrical quantities depend on a certain parameter which can be controlled. In this way, we should tune the value of critical density by acting on the parameter itself. A feasible choice for such a graph is made.

\subsection{The Graph}

Consider a collection of $M$ 3-$d$ square lattices, each one of finite size $N$. Consider then all the sites sharing the same positions $i,j$, with $i,j$ nearest neighbors belonging to the whole set of $M$ lattices. Remove connections and rearrange links by performing a random permutation of them, so that sites belonging to possibly different lattices may be connected as in Figure \ref{intrecci}, where a result a such a permutation in 1-$d$ is shown. Repeat this procedure for all bonds $\left\langle i,j \right\rangle$.
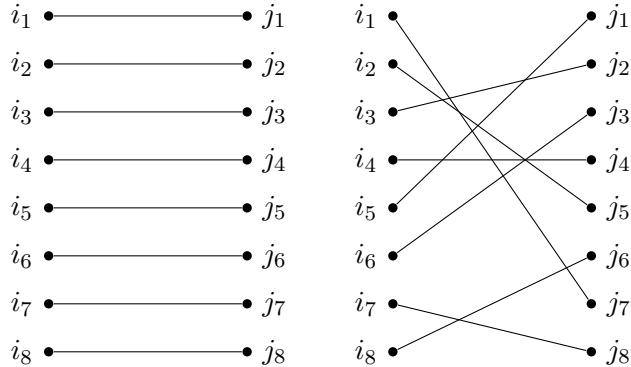
\begin{figure}[!htbp]
\centering
\begin{minipage}[c]{.2\textwidth}
\begin{tikzpicture}[
node distance=0.5cm, 
mydot/.style={
circle,
fill,
inner sep=1.25pt
}
]
\node[mydot,label={left:$i_1$}] (a1) {}; 
\node[mydot,below=of a1,label={left:$i_2$}] (a2) {}; 
\node[mydot,below=of a2,label={left:$i_3$}] (a3) {}; 
\node[mydot,below=of a3,label={left:$i_4$}] (a4) {}; 
\node[mydot,below=of a4,label={left:$i_5$}] (a5) {}; 
\node[mydot,below=of a5,label={left:$i_6$}] (a6) {}; 
\node[mydot,below=of a6,label={left:$i_7$}] (a7) {}; 
\node[mydot,below=of a7,label={left:$i_8$}] (a8) {}; 

\node[mydot,right=2.5cm of a1,label={right:$j_1$}] (b1) {}; 
\node[mydot,below=of b1,label={right:$j_2$}] (b2) {}; 
\node[mydot,below=of b2,label={right:$j_3$}] (b3) {}; 
\node[mydot,below=of b3,label={right:$j_4$}] (b4) {}; 
\node[mydot,below=of b4,label={right:$j_5$}] (b5) {}; 
\node[mydot,below=of b5,label={right:$j_6$}] (b6) {}; 
\node[mydot,below=of b6,label={right:$j_7$}] (b7) {}; 
\node[mydot,below=of b7,label={right:$j_8$}] (b8) {}; 

\path[-] (a1) edge (b1);
\path[-] (a2) edge (b2);
\path[-] (a3) edge (b3);
\path[-] (a4) edge (b4); 
\path[-] (a5) edge (b5);
\path[-] (a6) edge (b6);
\path[-] (a7) edge (b7);
\path[-] (a8) edge (b8); 
\end{tikzpicture}
\end{minipage}
\hspace{10mm}
\begin{minipage}[c]{.2\textwidth}
\begin{tikzpicture}[
node distance=0.5cm,
mydot/.style={
circle,
fill,
inner sep=1.25pt
}
]
\node[mydot,label={left:$i_1$}] (a1) {}; 
\node[mydot,below=of a1,label={left:$i_2$}] (a2) {}; 
\node[mydot,below=of a2,label={left:$i_3$}] (a3) {}; 
\node[mydot,below=of a3,label={left:$i_4$}] (a4) {}; 
\node[mydot,below=of a4,label={left:$i_5$}] (a5) {}; 
\node[mydot,below=of a5,label={left:$i_6$}] (a6) {}; 
\node[mydot,below=of a6,label={left:$i_7$}] (a7) {}; 
\node[mydot,below=of a7,label={left:$i_8$}] (a8) {}; 

\node[mydot,right=2.5cm of a1,label={right:$j_1$}] (b1) {}; 
\node[mydot,below=of b1,label={right:$j_2$}] (b2) {}; 
\node[mydot,below=of b2,label={right:$j_3$}] (b3) {}; 
\node[mydot,below=of b3,label={right:$j_4$}] (b4) {}; 
\node[mydot,below=of b4,label={right:$j_5$}] (b5) {}; 
\node[mydot,below=of b5,label={right:$j_6$}] (b6) {}; 
\node[mydot,below=of b6,label={right:$j_7$}] (b7) {}; 
\node[mydot,below=of b7,label={right:$j_8$}] (b8) {}; 

\path[-] (a1) edge (b7);
\path[-] (a2) edge (b5);
\path[-] (a3) edge (b2);
\path[-] (a4) edge (b4); 
\path[-] (a5) edge (b1);
\path[-] (a6) edge (b3);
\path[-] (a7) edge (b8);
\path[-] (a8) edge (b6); 
\end{tikzpicture}
\end{minipage}
\caption{(Right) example of random permutation of nearest neighbor links when $M=8$. (Left) disposition before the permutation.}\label{intrecci}
\end{figure}
\newline
In such a layered random graph the number of closed paths increases with the number of layers $M$, so that we conjecture the critical point to increase in a non trivial way with $1/M$, so that as $N\to \infty$:
\begin{equation}
\rho_c\left(M\right) = \rho_{\infty} + \mathrm{\Delta}\rho\left({1\over M}\right).
\end{equation}
Clearly when $M=1$ we recover the case of the simple cubic lattice, while for $M\to \infty$ we expect the critical density $\rho_{\infty}$ be the same as in the Bethe lattice. The conjecture even regards the possibly existence of a truly critical point $\rho_c\left(M\right)$ for $M>1$.
\newline
Even if an analytical frame concerning KA dynamics on layered random graphs is lacking, we here report some numerical results for several values of $M$ with the aim of supporting the underlying conjecture. 

\subsection{Numerical Results for Layered Graphs}

We run KA dynamics on layered graphs for $M=2$, 4, 8, 16 keeping the volume of each square lattice fixed to $N=14$, in order to prevent a mixing of finite size effects. As before, we registered for each $M$ the dynamical behavior of $\textsf{C}\left(t\right)$ and $\chi_4\left(t\right)$ from which we extracted decorrelation times. In such a way we get numerical estimates for $\rho_c\left(M\right)$.


\begin{figure}[!htbp]
\centering
\begin{tikzpicture}
\begin{axis}[
width=8cm,
height=5cm,
scale only axis,
ylabel=$\textsf{C}\left(t\right)$,
xlabel=$t$,
xmin=0.01,
xmax=1000000,
ymin=0,
ymax=1,
xmode=log,
log basis x={10},
xtick={0.0001,0.01,1,100,10000,1000000},
legend style={draw=black,thin,at={(0.03,0.4)},anchor=north west,legend cell align=left,font=\scriptsize},
]
\addplot+[only marks,mark size=0.35mm,mark options={line width=0.025pt,fill=gray},color=black]
table{chi2layer16_83.dat};
\addlegendentry{$M=16$};
\addplot+[only marks,mark size=0.35mm,mark options={line width=0.025pt,fill=gray},color=black]
table{chi2layer8_83.dat};
\addlegendentry{$M=8$};
\addplot+[only marks,mark size=0.35mm,mark options={line width=0.025pt,fill=gray},color=black]
table{chi2layer4_83.dat};
\addlegendentry{$M=4$};
\addplot+[only marks,mark size=0.35mm,mark options={line width=0.025pt,fill=gray},color=black]
table{chi2layer2_83.dat};
\addlegendentry{$M=2$};
\end{axis}
\end{tikzpicture}
\caption{Behavior of $\textsf{C}\left(t\right)$ at fixed $\rho=0.83$. The correlation times increases with $M$.}\label{obslayert}
\end{figure}
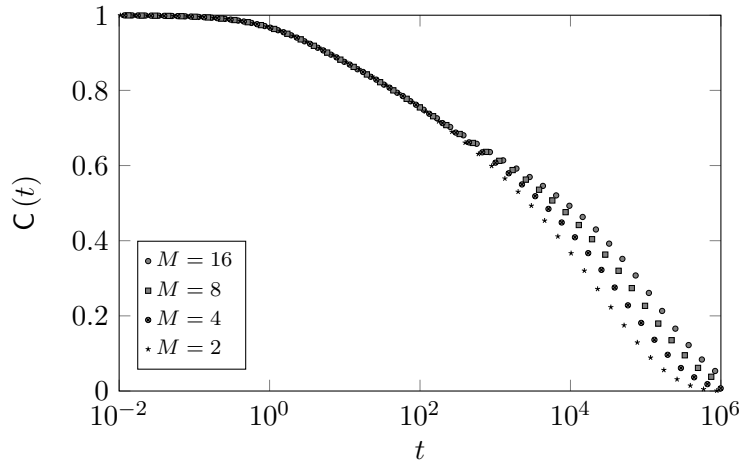


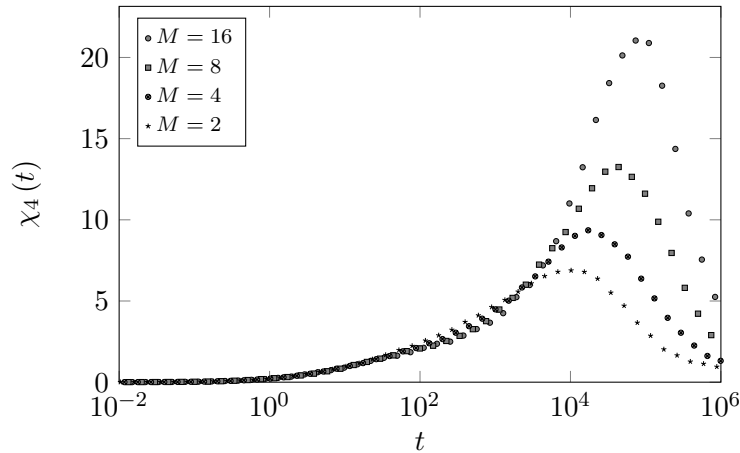
\begin{figure}[!htbp]
\centering
\begin{tikzpicture}
\begin{axis}[
width=8cm,
height=5cm,
scale only axis,
ylabel=$\chi_4\left(t\right)$,
xlabel=$t$,
xmin=0.01,
xmax=1000000,
ymin=0,
xmode=log,
log basis x={10},
xtick={0.0001,0.01,1,100,10000,1000000},
legend style={draw=black,thin,at={(0.03,0.97)},anchor=north west,legend cell align=left,font=\scriptsize},
]
\addplot+[only marks,mark size=0.35mm,mark options={line width=0.025pt,fill=gray},color=black]
table{chi4layer16_83.dat};
\addlegendentry{$M=16$};
\addplot+[only marks,mark size=0.35mm,mark options={line width=0.025pt,fill=gray},color=black]
table{chi4layer8_83.dat};
\addlegendentry{$M=8$};
\addplot+[only marks,mark size=0.35mm,mark options={line width=0.025pt,fill=gray},color=black]
table{chi4layer4_83.dat};
\addlegendentry{$M=4$};
\addplot+[only marks,mark size=0.35mm,mark options={line width=0.025pt,fill=gray},color=black]
table{chi4layer2_83.dat};
\addlegendentry{$M=2$};
\end{axis}
\end{tikzpicture}
\caption{Dynamical susceptibility at fixed $\rho=0.83$. Notice that the correlation time $t_4$ grows as $M$ increases as well as the size of correlated particles clusters.}\label{obslayer}
\end{figure}


\noindent
In Figure \ref{corrtimes} we show the behavior of $t^*\left(\rho\right)$ and $t_4\left(\rho\right)$ for the mentioned numbers of layers together with the simple cubic and the random regular graph cases. Notice that, as expected, the critical density decreases with $M$. By conjecturing $\rho_c\left(M\right)$ to be a truly transition point for the KA dynamics, we fitted data with a power law in order to get numerical estimates for $\rho_c\left(M\right)$. The behavior as a function of $1/M$ is shown in Figure \ref{fin}, together with a power law fit.
\newline
We do not rule out the possibility $\rho_c\left(M\right)$ to increase linearly with $1/M$. Indeed, finite size effects may be taken into account in the $1/M$ correction. They in principle may depend on $N$ and $M$ rather than $N\times M$, while we kept $N$ fixed in the numerical simulations. Clearly, this is not enough to prove our conjecture; however, $\rho_c\left(M\right)$ monotonically increases with the number of layers, which is the minimum required. 


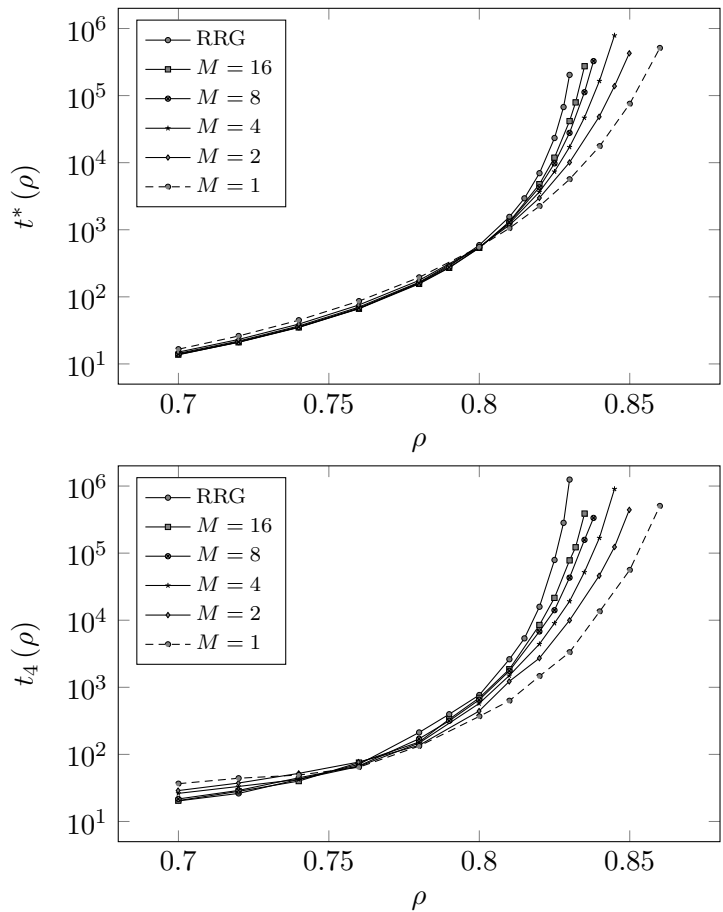
\begin{figure}[!htbp]
\centering
\begin{tikzpicture}
\begin{axis}[
width=8cm,
height=5cm,
scale only axis,
ylabel=$t^*\left(\rho\right)$,
xlabel=$\rho$,
ymin=5,
ymax=2000000,
xmin=0.68,
xmax=0.88,
ymode=log,
log basis y={10},
xtick={0.70,0.75,0.80,0.85},
ytick={10,100,1000,10000,100000,1000000},
legend style={draw=black,thin,at={(0.03,0.97)},anchor=north west,legend cell align=left,font=\scriptsize},
]
\addplot+[mark size=0.35mm,mark options={line width=0.025pt,fill=gray},color=black]
table{tauBL.dat};
\addlegendentry{RRG};
\addplot+[mark size=0.35mm,mark options={line width=0.025pt,fill=gray},color=black]
table{tau_M16.dat};
\addlegendentry{$M=16$};
\addplot+[mark size=0.35mm,mark options={line width=0.025pt,fill=gray},color=black]
table{tau_M8.dat};
\addlegendentry{$M=8$};
\addplot+[mark size=0.35mm,mark options={line width=0.025pt,fill=gray},color=black]
table{tau_M4.dat};
\addlegendentry{$M=4$};
\addplot+[mark size=0.35mm,mark options={line width=0.025pt,fill=gray},color=black]
table{tau_M2.dat};
\addlegendentry{$M=2$};
\addplot+[mark size=0.35mm,mark options={line width=0.025pt,fill=gray},color=black]
table{tauKA.dat};
\addlegendentry{$M=1$};
\end{axis}
\end{tikzpicture}

\begin{tikzpicture}
\begin{axis}[
width=8cm,
height=5cm,
scale only axis,
ylabel=$t_4\left(\rho\right)$,
xlabel=$\rho$,
ymin=5,
ymax=2000000,
xmin=0.68,
xmax=0.88,
ymode=log,
log basis y={10},
xtick={0.70,0.75,0.80,0.85},
ytick={10,100,1000,10000,100000,1000000},
legend style={draw=black,thin,at={(0.03,0.97)},anchor=north west,legend cell align=left,font=\scriptsize},
]
\addplot+[mark size=0.35mm,mark options={line width=0.025pt,fill=gray},color=black]
table{t_BL.dat};
\addlegendentry{RRG};
\addplot+[mark size=0.35mm,mark options={line width=0.025pt,fill=gray},color=black]
table{t_M16.dat};
\addlegendentry{$M=16$};
\addplot+[mark size=0.35mm,mark options={line width=0.025pt,fill=gray},color=black]
table{t_M8.dat};
\addlegendentry{$M=8$};
\addplot+[mark size=0.35mm,mark options={line width=0.025pt,fill=gray},color=black]
table{t_M4.dat};
\addlegendentry{$M=4$};
\addplot+[mark size=0.35mm,mark options={line width=0.025pt,fill=gray},color=black]
table{t_M2.dat};
\addlegendentry{$M=2$};
\addplot+[mark size=0.35mm,mark options={line width=0.025pt,fill=gray},color=black]
table{t_KA.dat};
\addlegendentry{$M=1$};
\end{axis}
\end{tikzpicture}
\caption{Decorrelation times $t^*\left(\rho\right)$, $t_4\left(\rho\right)$ for $M=$ 1, 2, 4, 8, 16 and for a random regular graph.}\label{corrtimes}
\end{figure}


\begin{figure}[!htbp]
\centering
\begin{tikzpicture}
\begin{axis}[
width=8cm,
height=5cm,
scale only axis,
ylabel=$\rho_c\left(M\right)$,
xlabel=${1/M}$,
xmin=0,
xmax=1,
legend style={draw=black,thin,at={(0.03,0.97)},anchor=north west,legend cell align=left,font=\scriptsize},
]
\addplot+[only marks,mark size=0.35mm,mark options={line width=0.025pt,fill=gray},color=black]
table{rho_tau.dat};
\addplot+[only marks,mark size=0.35mm,mark options={line width=0.025pt,fill=gray},color=black]
table{rho_t.dat};
\addplot[dashed,domain=0:1,samples=200]
{0.833731+0.0464829*x^0.625044};
\addplot[domain=0:1,samples=200]
{0.83327+0.0480348*x^0.504223};
\end{axis}
\end{tikzpicture}
\caption{The critical density as function of $1/M$ obtained by fitting decorrelation times $t^*\left(\rho\right)$ (square marks) and $t_4\left(\rho\right)$ (circle marks). Apart from the numerical discrepancy, $\rho_c\left(M\right)$ increases monotonically with $1/M$. The solid and the dashed lines refer to two different power law fits.}\label{fin}
\end{figure}
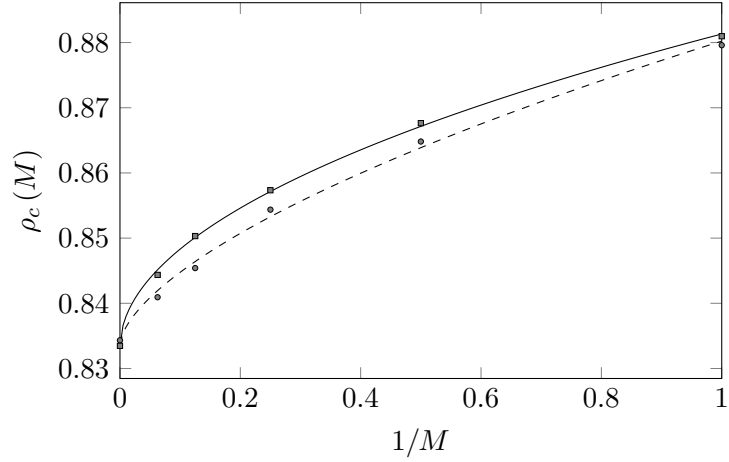


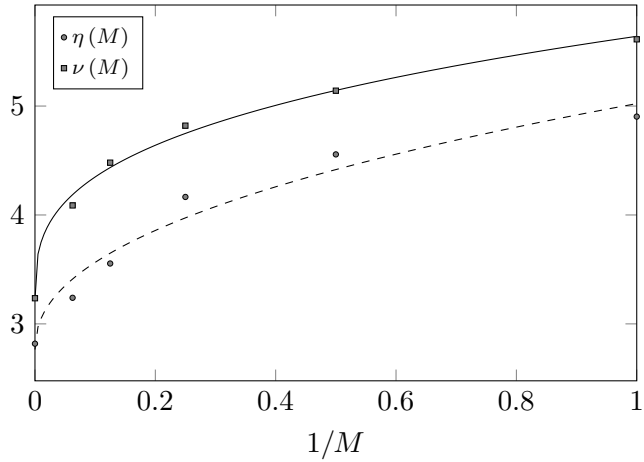
\begin{figure}[!htbp]
\centering
\begin{tikzpicture}
\begin{axis}[
width=8cm,
height=5cm,
scale only axis,
ylabel=$\textcolor{white}{\rho_c\left(M\right)}$,
xlabel=${1/M}$,
xmin=0,
xmax=1,
legend style={draw=black,thin,at={(0.03,0.97)},anchor=north west,legend cell align=left,font=\scriptsize},
]
\addplot+[only marks,mark size=0.35mm,mark options={line width=0.025pt,fill=gray},color=black]
table{gamma_tau.dat};
\addlegendentry{$\eta\left(M\right)$};
\addplot+[only marks,mark size=0.35mm,mark options={line width=0.025pt,fill=gray},color=black]
table{eta_t.dat};
\addlegendentry{$\nu\left(M\right)$};
\addplot[dashed,domain=0:1,samples=200]
{2.76483+2.25723*x^0.450774};
\addplot[domain=0:1,samples=200]
{3.22472+2.41517*x^0.332218};
\end{axis}
\end{tikzpicture}
\caption{Critical exponents $\eta\left(M\right)$, $\nu\left(M\right)$ at varying $M$ with the corresponding power law interpolations.}\label{criti}
\end{figure}


\section{Conclusions}

We introduced a multispin coding based algorithm with the purpose of speeding up numerical simulations for the Kob-Andersen model, which is one of the most powerful stochastic models which mimic cage effects in the real glasses. We tested the algorithm by running KA dynamics on a cubic lattice and on a random regular graph, for which analytical and numerical results do already exist. We then moved on an interpolating layered lattice we introduced in order to suitably tune particle entrapment. We conjectured the critical point which discriminates between an ergodic and a non-ergodic phase to exist in the thermodynamic limit and then we gave estimates at varying $M$.
\newline
Nevertheless, our analysis in this more general frame was purely numerical. In particular, a proof of the actual existence of such a transition at $M>1$ is lacking; moreover, despite the fact that the density at which a dynamical arrest takes place decreases with the typical number of inner closed paths, we did not establish a precise connection with $M$.

\section*{Acknowledgements}

I am grateful to Giorgio Parisi for enlightening remarks and for his continuous support.

\end{document}